\documentclass[aps,prd,superscriptaddress]{revtex4}
\usepackage{graphicx}
\usepackage{amssymb}
\usepackage{psfrag}

\begin{document}

\title{Universal correlations of trapped one-dimensional impenetrable bosons}

\author{D.M.~Gangardt}
\email[e-mail: ]{gangardt@lptms.u-psud.fr}
\affiliation{\mbox{Laboratoire de Physique Th\'eorique et Mod\`eles
  Statistiques, Universit\'e Paris Sud, 91405 Orsay Cedex, France}}
\affiliation{\mbox{Laboratoire Kastler-Brossel, Ecole Normale Sup\'erieure, 
24 rue Lhomond, 75231, Cedex 05, Paris, France}}

\date{\today}

\begin{abstract}
We calculate the asymptotic behaviour of the one body density
matrix of one-dimensional impenetrable bosons in finite size geometries. 
Our approach is based on a modification of the Replica Method from the
theory of disordered systems.  We  obtain explicit expressions
for  oscillating terms, similar to fermionic  Friedel
oscillations. These terms  are  universal and originate from the
strong short-range correlations between bosons in one dimension.
\end{abstract}

\maketitle

\section{Introduction}
Recently, one-dimensional (1D) Bose gases have been created in long
cylindrical traps by tightly confining the transverse motion of particles to
zero-point oscillations \cite{Experiments1D}. These experiments revived an
interest in exactly solvable one dimensional models of statistical
physics, in particular in 
the Lieb-Liniger model \cite{LiebLiniger1963} 
for one-dimensional bosons  interacting via a delta-function potential. 
In this case the  Bethe Ansatz solution accounts for  the ground 
state properties,  spectrum of elementary excitations
\cite{Lieb1963} and thermodynamics \cite{Yang1969}. 
In contrast to these properties, following from the  solution of relatively 
simple integral equations, the correlation functions are 
not easily  obtained  from the Bethe Ansatz  due to an
extremely complicated form  of the wave functions. In the  limiting cases of 
weak and strong  interactions  closed analytical results can  be 
found as perturbative expansions \cite{JKC,KorepinBook}. 

Another drawback of the Bethe Ansatz is the requirement  of  periodic
boundary conditions and thermodynamic limit. 
This  precludes the study of finite size effects,
except for a number of cases in which a  special symmetry of the confining 
potential \cite{Gaudin} is available. In most cases 
\cite{Theory,Gangardt2003}  the  
finite geometry was treated relying on the 
local density approximation, where the  macroscopic length
scale induced by the confining potential  is assumed to be well separated
from the  microscopic correlation length emerging from 
the Lieb-Liniger solution. 

A special case that  allows one  to go beyond the local density 
approximation, is the limiting case of an infinite coupling constant. 
Then the interactions are taken into account by mapping the system onto free
fermions as shown by Girardeau \cite{Girardeau60-65}. 
The resulting  system of impenetrable bosons, or 
``Tonks gas''  shares many of its properties with free fermions.
In the  present stage,  achieving  this strong-coupling  Tonks-Girardeau 
regime  is one of the  main
experimental  goals \cite{Tolra2003,Reichel2003,Paredes2004}. 
In this respect, it is  desirable to have results for finite systems  beyond
the local density approximation and to compare them directly
with experimental data.

From a theoretical point of view, the fermionic mapping 
leads to a considerable simplification of the general 
Bethe Ansatz expression for the wave functions. Historically, even 
before the Bethe Ansatz solution became available, Girardeau 
 \cite{Girardeau60-65} was the first to note the boson-fermion
correspondence.  He found a  simple expression for the ground state wave 
function of $N$ bosons in arbitrary potential $V(x)$
in the form of an absolute  value of the 
fermionic Slater determinant:
\begin{equation}
  \label{eq:gs}
  \Phi(x_1,\ldots, x_N) = \Big|{\det}_{k,l} [\varphi_{k} (x_l)]\Big|,
\end{equation}
where $\varphi_k (x)$ are one particle  wave functions   in the
potential $V(x)$. 

The equivalence of impenetrable bosons and free fermions can be 
stated as the equivalence of correlation functions. Indeed, any
correlation function of the density
is given  by the corresponding expression for fermions, since it is 
diagonal in field operators and does not involve phase correlations. The same
holds for the ground state energy and the spectrum of elementary excitations.
However, off-diagonal correlation functions of impenetrable
bosons 
are different from those of fermions, due to the
presence of the absolute value in (\ref{eq:gs}). This drastically changes 
phase correlations. The simplest and well studied
example of such off-diagonal correlation function 
is the  one-body density matrix 
\begin{equation}
  \label{eq:g1_def_0}
  g_1 (t,t') = N\int dx_2\ldots dx_N\, 
  \Phi^*(t,x_2,\ldots,x_N)\, \Phi(t',x_2,\ldots,x_N). 
\end{equation}
This quantity is of major importance for bosonic systems, since 
the eigenvalues of $g_1$  show the presence or absence of
Bose-Einstein condensation according to the criterion of 
Penrose and Onsager \cite{Penrose1956}.
In the translationally invariant case, $g_1$ depends only on the
relative distance $x=t-t'$ and its Fourier transform 
with respect to $x$ is the
momentum distribution of particles in the ground state. In this case
the condensation would manifest itself as a macroscopic occupation of 
the zero momentum state.  

The problem of calculating the one-body density matrix,
or equivalently the momentum distribution of impenetrable bosons,
has a long history in mathematical physics.  
First considered in 1963 by T.D.~Schultz \cite{Schultz1963}, 
the one-body density matrix was found in the form of a determinant 
with special properties, that is a
Toeplitz determinant\cite{GrenanderSzegoBook}.  
Using the known asymptotics of the  Toeplitz determinants
it was possible  to prove the absence of Bose-Einstein 
condensation by showing the  power-law decay  of the  one-body 
density matrix at large  distance. The precise form of this power 
law was obtained later by Lenard in  \cite{Lenard1964}.
His calculation resulted in the following long distance behaviour:  
\begin{equation}
  \label{eq:g1_smooth}
  \frac{\bar{g}_1 (x)}{n} =  \frac{\rho_\infty}{|k_F x|^{1/2}},
\end{equation}
where $n=k_F/\pi$ is the density of particles.    
In their \textit{tour de force}, Vaidya and Tracy \cite{VaidyaTracy1979}
calculated from the first principles the asymptotic long-distance 
behaviour of the one body density matrix (the short distance 
behaviour has also  been calculated)  and found the expression in 
(\ref{eq:g1_smooth}) as the  leading term with 
$\rho_\infty = \pi e^{1/2} 2^{-1/3} A^{-6}=0.92418\ldots$
where $A=\exp\left(1/12-\zeta'(-1)\right)=1.2824\ldots$ 
is Glaisher's constant, related  to the Riemann zeta  function $\zeta(z)$.
They also succeeded in obtaining the sub-leading terms. 
This work has been extended to higher order terms by Jimbo \textit{et
  al.} \cite{Jimbo1980} who related $g_1(x)$ to the solution of a certain
non-linear differential equation, the Painlev\'e equation of the  fifth kind. 
The general  structure of the large-distance expansion of the one-body density
matrix consists of trigonometric
functions, cosines or sines of multiples of $2k_F x$, each
such trigonometric term is  multiplied by 
series of even or odd powers of $1/k_F x$, respectively.
Using a hydrodynamic approach, Haldane \cite{Haldane1981} has shown that 
this structure is  a general property of  any one-dimensional
compressible  liquid, bosonic or fermionic. 
This method is unable to predict exact coefficients of the $2k_F x$
harmonics, which should be calculated using exact methods. For
example, according to \cite{VaidyaTracy1979,Jimbo1980} the first oscillatory 
correction to Eq.~(\ref{eq:g1_smooth}) is given by  
\begin{eqnarray}
  \label{eq:vaidya_tracy}
  \frac{g_1 (x)}{\bar{g}_1 (x)} -1 &=& +
    \frac{1}{8}\frac{\cos 2k_F x}{(k_Fx)^2}. 
\end{eqnarray}
Together with $\rho_\infty$, the coefficients of sub-leading
oscillatory terms provide the full information on the  long distance
asymptotics of the one body density matrix. It is then 
desirable to have  these coefficients in a simple analytical form or
to be able to calculate them by a perturbation theory.

Experimental conditions for  obtaining  the Tonks-Girardeau regime 
require a small number of particles, which in an isolated
system can be as  small as $N\sim 100$ atoms \cite{Reichel2003}. 
It is then important to
be able to extend the results of Eqs.~(\ref{eq:g1_smooth}) and
(\ref{eq:vaidya_tracy}) to finite size geometries. 
The case of harmonic confinement is directly related to 
experiments. It  was recently studied analytically in \cite{Gangardt2003} by
using Haldane's hydrodynamic approach and, independently, in
\cite{Forrester2003} by  the Coulomb gas analogy (for numerical
results, see the work \cite{Papenbrock2003} and references therein). The
expression for the  leading term in the one-body density matrix was obtained,
generalising the expression in Eq.~(\ref{eq:g1_smooth}) to a non-uniform 
density profile.   The case of a circular geometry has
been considered by Lenard \cite{Lenard1972} who conjectured
the main smooth contribution, analogous to (\ref{eq:g1_smooth}) in
the form:
\begin{equation}
  \label{eq:circ_smooth}
  \bar{g}_1 (\alpha) = \frac{N \rho_\infty}{|N\sin\pi\alpha|^{1/2}},
\end{equation}
where $2\pi\alpha$ is the angle between two points on the circle.
This result was justified  rigorously by Widom \cite{Widom1973} using the 
theory of Toeplitz determinants. It is worth mentioning here that
this result follows straightforwardly from the conformal field theory
(see e.g.\cite{TsvelikBook} ), which to some extent is equivalent to
the hydrodynamic \cite{Haldane1981} and Coulomb gas
\cite{Forrester2003,Forrester2003_Dirichlet} approaches. 
These methods are capable to predict long-wavelength, 
large scale behaviour of correlation functions, while giving only
qualitative answers for details on a scale of mean inter-particle distance.  
In both geometries, harmonic and circular, the finite size
corrections, analogous to the expression in Eq.~(\ref{eq:vaidya_tracy})  
remain unknown.

Here we present  the first  calculation of  finite
size  corrections to the one-body density matrix 
for both harmonic confinement and  circular geometry. 
We compare our analytical expressions
with numerical calculations based on the exact representation of 
the one-body density matrix as a
Toeplitz determinant and find excellent agreement.  
We also take on the task of reproducing Vaidya and
Tracy asymptotic expansion in the thermodynamic limit using both
systems as a starting point and increasing the number of particle and
the system size in a way to preserve a constant density. 
Our calculations are the first to  resolve analytically the sign ambiguity
of the oscillating terms, similar to that in
Eq.~(\ref{eq:vaidya_tracy}) which appears in the studies of Painlev\'e
representation for  the one-body density matrix \cite{Jimbo1980,Creamer1981}. 
We find that the sign of all oscillating terms, such as
(\ref{eq:vaidya_tracy})  in the Vaidya and Tracy
expansion \cite{VaidyaTracy1979} should  be reversed \cite{rmrkDunjko}.

Our calculations are based on a modification of  the Replica
Method developed in the theory of disordered systems
\cite{Edwards1975} and  applied recently to  random matrices
\cite{Verbaarschot1985,KamenevMezard1999,Yurkevich1999} and
Calogero-Sutherland models \cite{GangardtKamenev2001}. Recent progress
\cite{Kanzieper2002,Splittorf2003} in calculations based on the
Replica Method has elucidated its  intimate relation  
to the theory of non-linear integrable systems. The present work gives
yet another example of this interconnection.

The paper is organised as follows. In  Section~\ref{sec:model} 
we  present our method and illustrate it by calculating  the  
amplitude  of removing a particle from the ground
state of harmonically trapped impenetrable bosons.
The one-body density matrix for harmonically confined bosons is
calculated in  Section~\ref{sec:corr}. 
Section~\ref{sec:circular_geometry} describes the calculation of the
one body density matrix in the circular geometry. We calculate
sub-leading terms  using perturbation theory in
Section~\ref{sec:perturb}.  The conclusions are
presented in Section~\ref{sec:disc} and mathematical details are
given in two Appendices.

\section{\label{sec:model} Description of the method}
The task of reproducing and  extending
the original calculations of Vaidya and Tracy to finite systems
 is obscured  by the technical complexity
of their method which consist in the asymptotic expansion of
a Toeplitz determinant assuming  the size of the determinant tending to 
infinity in order to reproduce the continuous limit. 
It is therefore important to have 
an alternative way of representing the one-body density matrix. 
One starts with  the  representation  of the  one-body density matrix 
as an $N$-dimensional integral , first proposed by  Lenard \cite{Lenard1964}: 
\begin{equation}
  \label{eq:circ_g1_def}
  g_1(\alpha) = \frac{1}{N!}  \int_{0}^1 d^N \theta \left|\Delta_N
    \left(e^{2\pi i \theta_1},\ldots,e^{2\pi i \theta_N}\right)\right|^2 
  \prod_{l=1}^N
  \left|1-e^{2\pi i \theta_l}\right| \left|e^{2 \pi i \alpha}-e^{2\pi
  i \theta_l}\right| \equiv \left\langle \prod_{l=1}^N \left|1-z_l\right|
    \left|e^{2 \pi i \alpha}-z_l\right|\right\rangle 
\end{equation}
which follows immediately from the definition (\ref{eq:g1_def_0}). We
have chosen here a  circular geometry of $N+1$ particles described by
cyclic coordinates $0<2\pi\theta_l<2\pi$ with periodic boundary
conditions. The ground state wave function of $N+1$ particles
is given by the absolute value of the  Slater determinant (\ref{eq:gs}) 
composed of plane waves 
$\varphi_{k+1} (z_l) = \exp(2\pi i k\theta_l)=z^k_l$. It is then   
identified  with the absolute value of the Vandermonde determinant
\begin{equation}
  \label{eq:vandermonde}
  {\det}_{1\le k,l\le N+1} [\varphi_{k} (z_l)] = 
  \Delta_{N+1}(z)=\Delta (z_1,\ldots,z_{N+1}) =
  \prod_{1\le j<k\le N+1} (z_i-z_j).
\end{equation}
The above expression is factorised straightforwardly to yield the
expression being integrated in (\ref{eq:circ_g1_def}). Consider now 
a positive integer $n$ and correlation function
\begin{equation}
  \label{eq:average}
  Z_{2n} (\alpha) = \left\langle \prod_{l=1}^N \left(1-z_l\right)^{2n}
    \left(e^{2 \pi i \alpha}-z_l\right)^{2n}\right\rangle ,
\end{equation}
Our method is based on the fact that the
one-body density matrix (\ref{eq:circ_g1_def}) can be obtained from
$Z_{2n}$ by suitable analytical continuation to $n=1/2$. It happens
that $Z_{2n}$ can be evaluated straightforwardly in the asymptotic
large $N$ limit. In this respect the expression (\ref{eq:average}) 
supplemented with proper procedure for  analytic continuation 
$n\to 1/2$ is the desired  alternative representation of the one-body
density matrix.

The idea of
calculating the averages of the absolute value of a non-positive
definite function was put
forward by Kurchan \cite{Kurchan1991} and was named a modification of
the Replica trick. In the present context replica means the following.   
In 
\cite{Verbaarschot1985,KamenevMezard1999,Yurkevich1999,GangardtKamenev2001}  
the correlation function (\ref{eq:average}) was expressed using a dual  
representation involving an integral over 
components of a $n$-dimensional field,  with the action symmetric
under  rotations in the space of the components.   
This zero-dimensional field theory was considered in the $n\to 0$ limit to
obtain  density-density correlation function. In the present work
a different limit $n\to 1/2$ is taken to
reproduce the off-diagonal correlation function
(\ref{eq:circ_g1_def}).  However the idea of  the analytical
continuation in  $n$ is common to the above mentioned works and 
we use heavily the techniques introduced in \cite{KamenevMezard1999}. 
For example, we use the same dual representation of (\ref{eq:average}) and
evaluate it  in the asymptotic large $N$ limit.   

For $n$ integer, the result consists of a main smooth contribution 
and exactly $n$ oscillatory
corrections. This is an expected structure of one-dimensional
correlation functions conjectured by Haldane \cite{Haldane1981} form
his hydrodynamical approach. 
The sensible analytic continuation  $n\to 1/2$ is then performed 
in a way  to preserve this structure and the resulting expression
is believed to represent the large $N$ limit of the (unknown)
analytical continuation in $n$. The lack of rigour in this  approach
is shared by most replica calculations and is  justified \textit{a
  posteriori} by
remarkably transparent resulting expressions which are in full
agreement with the numerics as well as the known analytical results.

To demonstrate the method in detail and set up notations we first
calculate the ground state amplitude $A(t)$ of impenetrable bosons in
a harmonic potential:
\begin{equation}
  \label{eq:gs_ampldef1}
  A(t) ={}_N\langle \Psi(t)\rangle_{N+1} 
\end{equation}
This quantity describes the  probability amplitude to remove a particle
at position $t$ by acting with annihilation operator $\Psi(t)$ 
from the ground state of $N+1$ particles and leave the
system in the ground state of $N$ particles. It  can be considered as a
many-body wave function of the removed particle. We
consider a geometry different from circular, since for the latter the
ground state amplitude is just square root of mean density
independently of the position due to the translational invariance.  
We consider the system confined by harmonic potential 
$V(x)=m\omega^2x^2/2$ for which $A(t)$ has a non-trivial position
dependence. The one particle orbitals are given by eigenfunctions of
harmonic oscillator:
\begin{equation}
  \label{eq:orbitals}
  \phi_m (x) = \frac{1}{\sqrt{c_m}} H_m
  \left(x\sqrt{\frac{N}{2}}\right) e^{-\frac{N}{4} x^2},
  \;\;\;\;c_m=\left(\frac{2\pi}{N}\right)^\frac{1}{2} 2^m m!  ,
\end{equation}
and  $H_m$ are Hermite polynomials. We measure the
distances in units of half of the Thomas-Fermi radius $R=\sqrt{2\hbar
N/m\omega}$, corresponding to  half the size of the particles cloud in the
large $N$ limit. The  ground state wave function is obtained similarly to
the uniform case taking the  absolute value of the  fermionic Slater 
determinant:
\begin{equation}
  \label{eq:gs_h}
  \Phi(x_1,\ldots, x_N) = \Big|{\det}_{l,m} [\phi_{m-1} (x_l)]\Big|=
\frac{1}{\sqrt{S_N(N)}} \Big|\Delta_N(x)\Big| e^{-\frac{N}{4}\sum x_j^2}.
\end{equation}
The last identity follows from the 
linearity of the  determinant with respect to its columns, which
enables us to write the determinant of Hermite polynomials $H_m(x)$ as a
determinant of  their leading monomials $H_m (x)\sim x^m$.
The normalisation constant is given 
by Selberg integral \cite{MehtaRandMatr,ForresterWeb} of Hermite type
\begin{equation}
  \label{eq:selberg}
  S_N (\lambda) = \int_{-\infty}^\infty d^N x\, \Delta_N^2 (x)\,
  e^{-\frac{\lambda}{2}\sum x_j^2} = \lambda^{-N^2/2} (2\pi)^{N/2}
  \prod_{j=1}^N \Gamma(1+j) .
\end{equation}
Using the definition (\ref{eq:gs_ampldef1}) of the ground state 
amplitude leads to the expression
\begin{equation}
  \label{eq:gs_ampl_def}
  A(t) =
  \frac{e^{-\frac{N}{4}t^2}}{\sqrt{S_N (N) S_{N+1}(N) }}
  \int_{-\infty}^\infty d^N\! y\, \Delta_N^2 (y) e^{-\frac{N}{2}\sum
  y_j^2} \prod_{j=1}^N |t-y_j| 
\end{equation}
To deal with absolute value in this expression we use the identity
\begin{equation}
  \label{eq:repr}
|t-y_j|^{2n} = (t-y_j)^{2n},
\end{equation}
valid for integer $n$. Performing the  analytical continuation 
$n\to 1/2$ in the end of the  calculations, one recovers the expression
(\ref{eq:gs_ampl_def}). To treat the ground state amplitude and
one-body density matrix on the same footing
it is convenient to consider a more general quantity
\begin{equation}
  \label{eq:znnN}
  Z_{m}(t_1,\ldots,t_{m})= \frac{1}{S_N(N)}\int_{-\infty}^\infty
  d^N\! y\, \Delta_N^2 (y) e^{-\frac{N}{2}\sum y_j^2}
  \prod_{j=1}^N\prod_{a=1}^{m} (t_a-y_j)
\end{equation}
The variables $t_a$, different in general, are later taken equal to a single 
value $t$,
\begin{equation}
  \label{eq:zt_replica}
  Z(t)=\lim_{n\to 1/2} Z_{2n} (t)\equiv\lim_{n\to 1/2} Z_{2n}
  (\underbrace{t,\ldots,t}_{2n}) 
\end{equation}
to recover Eq.~(\ref{eq:gs_ampl_def}) up to a normalisation:
\begin{equation}
  \label{eq:gs_ampl2}
  A(t)= \sqrt{\frac{S_N
   (N)}{S_{N+1}(N)}}e^{-\frac{N}{4}t^2} Z(t).
\end{equation}

It is crucial that one cannot  set $2n=1$ directly  in (\ref{eq:znnN}),
which means that $Z (t)$ in (\ref{eq:zt_replica}) is different from $Z_1
(t)$.  The latter is nothing
but a fermionic ground state amplitude obtained by removing the
absolute value in (\ref{eq:gs_ampl_def}). It is given by the wave
function (\ref{eq:orbitals}) of $N+1$-th particle removed from the
system  and results in an expression that oscillates rapidly
around zero with period equal to the mean inter-particle separation.
In contrast, the bosonic ground state amplitude, obtained in the large $N$
limit, is expected to contain a smooth positive leading term which is
a  direct consequence of positivity of the ground state wave function 
(\ref{eq:gs_h}). Our claim is that it can be obtained from $Z(t)$. 

Recently the leading smooth contribution to 
$Z_{2n} (t)$  for integer $n$  was evaluated in the large $N$  
limit by Br\'ezin and Hikami  \cite{BrezinHikami2000}   and it was 
shown to survive the   analytical continuation  in $n$ off the integers. 
This suggests that we deal with two functions of variable $n$, 
one for fermions and one for bosons, or equivalently one function with
two branches, a  fermionic one and a bosonic one  
which coincide at integer $n$ but
become different as $n$ is moved away from integers. The goal of the
present calculation is the generalisation of  the bosonic analytic
continuation considered in \cite{BrezinHikami2000} to the whole
asymptotic expression for the  ground state amplitude.
 
To illustrate the ideas above  we start with a remarkable duality
transformation, which represents the  
$N$-fold integral $Z_{2n}$ in
Eqs.~(\ref{eq:znnN}),~(\ref{eq:zt_replica})
as an integral over $m=2n$ variables:
\begin{eqnarray}
  \label{eq:Zn_dual}
  Z_{2n} (t) = \frac{1}{S_{2n} (N)} \int_{-\infty}^\infty
  d^{2n}x\, \Delta^2_{2n} (x)
  e^{-N \sum_a S(x_a,t)}
\end{eqnarray}
with an effective action
\begin{equation}
  \label{eq:sxt}
  S(x,t)=\frac{(x-it)^2}{2} - \ln x +\frac{\pi i}{2}.
\end{equation}
This representation is exact and its proof can be found in the 
mathematical literature (see \cite{Duality} and references therein). 
For the case of harmonic
confinement the proof was presented in \cite{KamenevMezard1999}  
using the Random Matrix Theory. This proof has the  advantage  to be
readily extended to  deal with two-point correlation functions in 
harmonic potential. In Appendix~\ref{app:duality_der} we present yet 
another proof using second quantization for fermions.

The dual representation (\ref{eq:Zn_dual}) provides us with an
alternative representation of the $N$-fold integral~(\ref{eq:gs_ampl_def})
and is the starting point of the large $N$ asymptotic expansion.
The dominant contribution comes from
the saddle points of the action (\ref{eq:sxt}), situated at
\begin{eqnarray}
  \label{eq:xpm}
  x_\pm&=&\frac{it}{2}\pm\sqrt{1-\frac{t^2}{4}} = \pm e^{\pm i\phi},
  \;\;\;\;\;\sin \phi=\frac{t}{2}
\end{eqnarray}
for $t$ inside the ground state density support $-2<t<2$.  The
stationary value of the action and its second derivative at these
points read
\begin{eqnarray}
  S(x_\pm,t)&\equiv&S_\pm
=\frac{e^{\mp2i\phi}}{2}\mp i\phi \pm \frac{\pi i}{2},
  \label{eq:spm}
  \\ S''(x_\pm,t)&\equiv&S''_\pm=2 e^{\mp i\phi}\cos\phi
  \label{eq:s2pm}
\end{eqnarray}
To take into account all possible saddle points we put $l$ variables
$x_a$ in the vicinity of $x_-$ and $2n-l$ variables in the  vicinity of
$x_+$, such that
\begin{eqnarray}  
  x_a&=&x_-+\xi_a/\sqrt{N},\;\;\;\;a=1,\ldots,l \nonumber\\
  x_b&=&x_++\xi_b/\sqrt{N},\;\;\;\;b=l+1,\ldots,2n .
  \label{eq:x_exp}
\end{eqnarray}
In what follows the saddle points with $l=0$ or $2n$ will be referred
to as  \emph{replica symmetric} for obvious reasons. 
The other saddle points break the replica symmetry to some degree, the
maximal degree of replica symmetry breaking  happens for $l=n$.  
We shall see that the analytical continuation of the 
contribution of this saddle point leads to the result of 
\cite{BrezinHikami2000}. On the other hand,
in the calculation of the fermionic ground state amplitude, the  
replica symmetry is preserved, since in this case the contribution comes 
from the replica symmetric saddle points.

The  pre-exponential factor in (\ref{eq:Zn_dual}) 
given by Vandermonde determinant vanishes at each saddle point
(\ref{eq:x_exp}) and should be expanded as 
\begin{equation}
  \label{eq:vandermonde_saddle}
  \Delta^2_{2n} (x)
  =\left(\frac{1}{\sqrt{N}}\right)^{l(l-1)+(2n-l)(2n-l-1)}
  (x_--x_+)^{2l(2n-l)} \Delta^2_l (\xi_a)\Delta^2_{2n-l} (\xi_b)
\end{equation}
to yield a non-zero result. Using this expansion
the integration  of the fluctuations near each saddle point is performed by 
using the Selberg integral (\ref{eq:selberg}).  Combining the results
of integration with the main saddle point contribution and summing
over all saddle points we represent the original integral 
(\ref{eq:Zn_dual}) as a  sum of  $2n+1$ terms
\begin{equation}
  \label{eq:sum_saddle1}
  Z_{2n}(t) = \sum_{l=0}^{2n} F^l_{2n} N^{l(2n-l)} \frac{ (x_+-x_-)^{2l(2n-l)}}
  {\left(\sqrt{S''_-}\right)^{l^2}\left(\sqrt{S''_+}\right)^{(2n-l)^2}}
  e^{-NlS_--N(2n-l)S_+}.
\end{equation}
The factors
\begin{equation}
  F_{2n}^l = {2n \choose l} \frac{
    \prod_{a=1}^l\Gamma(a+1)\;\prod_{b=1}^{2n-l}\Gamma(b+1)} {
    \prod_{c=1}^{2n} \Gamma(c+1)} = \prod_{a=1}^l\frac{\Gamma(a)}
    {\Gamma(2n+1-a)}
  \label{eq:Anl}
\end{equation}
combine the numerical factors originating from Selberg integral
and the binomial coefficient equal to the number of ways to choose $l$
variables $x_a$ in the vicinity of $x_-$ out of total $2n$ variables.

As $|S_+|=|S_-|$, the $N$ dependence of each term in
(\ref{eq:sum_saddle1}) enters only through the factor $N^{l(2n-l)}$. Therefore
the central term  of the sum
(\ref{eq:sum_saddle1}) with  $l=n$ corresponding to the maximal degree
of the replica symmetry breaking  gives the dominant non-oscillating 
contribution to $Z_{2n} (t)$. The other terms, including the
replica symmetric terms $l=0,2n$  are at least 
$1/N$ times smaller and oscillate rapidly. This is to be contrasted with the
replica approach in \cite{KamenevMezard1999,GangardtKamenev2001}
where in the limit $n\to 0$ the dominant contribution comes from the 
replica symmetric points and the role of the saddle points with
broken replica symmetry is to provide oscillatory corrections.

Observing the behaviour of (\ref{eq:sum_saddle1}) for integer $n$ 
we are lead to the conjecture that the right
analytical continuation off the integer $n$ following the bosonic
branch would preserve this form: one smooth term plus oscillatory 
corrections. We now prepare the expression (\ref{eq:sum_saddle1}) for
the analytical continuation. To this end it is convenient to change 
the summation variable $l=k+n$  and
factorise the amplitude (\ref{eq:Anl}) as 
$F_{2n}^l = A_n D^{(n)}_k$, where the first factor is given by the product
\begin{equation}
  \label{eq:Ank}
  A_n=\prod_{a=1}^{n}\frac{\Gamma(a)} {\Gamma(2n+1-a)},
\end{equation}
where $n$ enters explicitly as the range of product. The analytical 
continuation of this factor to non-integer values of $n$ is described,
for example in \cite{Forrester2003,BrezinHikami2000}. We reproduce it    
in Appendix~\ref{app:analAn}, where it is shown that 
$\rho_\infty=A^2_{1/2}/\sqrt{2}$.  

The analytic continuation of the second factor 
\begin{equation}
  \label{eq:Dnk}
  D^{(n)}_{k}= \prod_{a=1}^{k}
  \frac{\Gamma\left(n+a\right)}
  {\Gamma\left(n+1-a\right)}= 
  \frac{\Gamma(n+k)}{\Gamma(n+1-k)} D^{(n)}_{k-1},\;\;\;\;\; k>0;
  \;\;\;\;D^{(n)}_0=1;\;\;\;\;D^{(n)}_{k} = D^{(n)}_{-k}
\end{equation}
to non-integer values  of $n$  is straightforward. 
For integer $n$  the coefficients $D^{(n)}_k$   vanish for  $|k|>n$ 
due to the divergence of the gamma function in the denominator, so for integer
$n$ the sum in (\ref{eq:sum_saddle1}) can be 
formally extended to $-\infty <k<+\infty$. For general values 
of $n$, the factor  $D^{(n)}_k$ has a  non-zero value for any $k$,
which results in genuine  infinite series. These series are asymptotic, rather
than convergent, but it can be shown by using the Stirling formula
that the coefficients decrease very fast for small values of $k$. 
Hence,  for large enough $N$, few terms around $k=0$ provide an 
excellent approximation for the sum. We rewrite the series 
(\ref{eq:sum_saddle1}) using the new summation variable:
\begin{equation}
  \label{eq:sum_saddle3}
  Z_{2n}(t) = A_n \left(2N\cos\phi\right)^{n^2} e^{-Nn\cos 2\phi} 
  \sum_{k=-\infty}^\infty \frac{D^{(n)}_k }{\left(8
    N \cos^3\phi\right)^{k^2}}\, e^{-i N k \Theta -2 i n k \phi},
\end{equation}
where $\Theta(t)=2\phi+\sin 2\phi+\pi=2\pi\int_{-2}^t \rho (s) ds$ is
expressed as an integral of  the mean density of particles
$\rho(t)=(1/\pi)\cos\phi$,
given by the celebrated Wigner semi-circle law:
\begin{equation}
  \label{eq:density}
  \rho(t)=\frac{1}{\pi}\sqrt{1-\frac{t^2}{4}}. 
\end{equation}
The result has  the expected form consisting of  sum of smooth and 
oscillating parts:
\begin{eqnarray}
  Z_{2n} (t) &=&A_n \left(2\pi N\right)^{n^2} e^{-Nn\left(1-t^2/2\right)}
\nonumber\\ &\times&\left[\rho(t)\right]^{n^2}
  \left[1+2\sum_{k=1}^\infty
  \frac{D^{(n)}_k}{\Big[8N\pi^3\rho^3(t)\Big]^{k^2}} \cos \left(
  kN \Theta(t)+2nk\phi(t)\right)\right] .
  \label{eq:Zn0_res}
\end{eqnarray}
This is the desired asymptotic representation of $Z_n$. For an integer
$n$ it has a simple general structure: a smooth main contribution plus
oscillatory corrections. This structure is preserved if analytical
continuation $n\to 1/2$ is applied term by term. The validity of 
the change of summation variable $l=n+k$ for a non-integer $n$ is
explained as follows: we assume that in
fact the infinite asymptotic series (\ref{eq:sum_saddle3}) is the correct
expression of the large $N$ asymptotics for the  analytical continuation of 
$Z_{2n} (t)$ to arbitrary  $n$. For integer $n$ the sum terminates and 
can be rewritten as (\ref{eq:sum_saddle1}) after the corresponding
change of the summation index. 

Multiplying
(\ref{eq:Zn0_res})  by the  normalisation factors defined in 
(\ref{eq:g1_def}) and using the asymptotic expansion
\begin{eqnarray}
  \label{eq:norm}
  \frac{S_N (N)}{S_{N+1} (N)}
  =\frac{1}{\sqrt{2\pi}}\frac{N^{N+\frac{1}{2}}}{(N+1)!}  \sim
  \frac{e^N}{2\pi N}.
\end{eqnarray}
we obtain the desired asymptotic expression for the ground state amplitude
\begin{eqnarray}
  A (t) &=& \sqrt{\rho_\infty} \left(\frac{\rho(t)}{\pi
      N}\right)^{1/4} \left[1+ 2\sum_{k=1}^\infty
    \frac{D^{(1/2)}_k}{\Big[8N\pi^3\rho^3(t)\Big]^{k^2}} \cos 
    \left[
      kN \Theta(t)+k\phi(t)\right] \right].\label{eq:gs_ampl_res1}\\ 
  D^{(1/2)}_1&=&1/2,\;\;D^{(1/2)}_2=-3/8,\ldots,\qquad
  D^{(1/2)}_{k+1}=
  D^{(1/2)}_k\frac{\Gamma\left(k+\frac{1}{2}\right)}
  {\Gamma\left(\frac{3}{2}-k\right)} 
\label{eq:gs_ampl_res}
\end{eqnarray}
The explicit values of the coefficients of oscillatory terms are obtained
from (\ref{eq:Dnk}).

In recent papers \cite{Kanzieper2002,Splittorf2003} the functions 
$Z_m(t)$ were shown to satisfy a remarkable recursion relation, a Toda
Lattice hierarchy, extensively studied in the theory of non-linear
integrable systems \cite{TeschlBook}. The same Toda Lattice equations
relates the solutions of Painlev\'e equations, where $m$ enters as a
parameter not restricted to integer values \cite{Forrester2001}. 
This observation was crucial  for \emph{exact} evaluation of the 
replica limit $m\to 0$ without relying on the large $N$ asymptotics.

In the present framework the quantity of interest is $Z_m(t)$ for 
$m\to 1$, so according to \cite{Forrester2001,Kanzieper2002} 
it can be related straightforwardly to  the solution of Painlev\'e IV 
equation. The boundary conditions, obtained from the small $t$
expansion of the ground state amplitude (\ref{eq:gs_ampl_def})
distinguishes between bosonic and fermionic branches of $Z_1 (t)$.
Our  analytic continuation with broken  replica symmetry chooses 
automatically the correct  boundary conditions for bosons. 

To justify numerically the result (\ref{eq:gs_ampl_res}) we have
calculated the ground state amplitude as a determinant
\begin{eqnarray}
  A(t)=\frac{N!\,e^{-Nt^2/4}}{\sqrt{S_N(N) S_{N+1}(N)}}\
  \det_{1\le j,k\le N} A_{j+k-2} (t),
  \label{eq:toeplitz_a}
\end{eqnarray}
where the matrix elements can be expressed by gamma function and 
incomplete gamma function $\gamma(\alpha,x)$ as
\begin{eqnarray*}
  \label{eq:f_inc_gamma}
  A_m (t)&=&\int_{-\infty}^{\infty} dy\,|t-y| y^m e^{-Ny^2/2} =
  \left(\frac{2}{N}\right)^{m/2+1}
  \left[f\left(m+1,\frac{Nt^2}{2}\right)-
    \sqrt{\frac{Nt^2}{2}}f\left(m,\frac{Nt^2}{2}\right)\right]\\
  f(m,x) &=& \frac{1-(-1)^m}{2}\Gamma\left(\frac{m+1}{2}\right) 
  -\gamma\left(\frac{m+1}{2},\frac{Nt^2}{2}\right).
\end{eqnarray*}
\begin{figure}[tbp]
  \centering
  \psfrag{a}{$(\pi/\rho_\infty)^{1/2} N^{1/4} A(t)$}
  \psfrag{t}{$t$}
  \includegraphics[width=14cm]{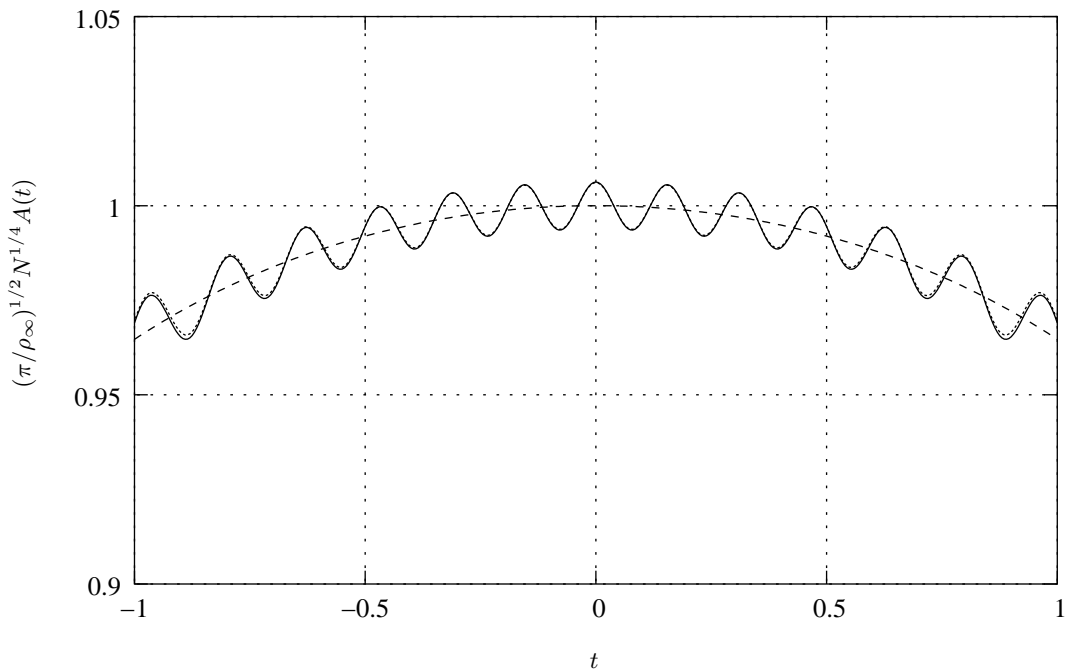}
  \caption{Comparison between asymptotic expression
    for the ground state amplitude (\ref{eq:gs_ampl_res})  (solid line) 
    of $N=20$ particles in harmonic potential evaluated up to the $k=1$ terms
    and exact results (dotted line) based on numerical 
    evaluation of the determinant (\ref{eq:toeplitz_a}). 
    The contribution given
    by the smooth term $k=0$ in (\ref{eq:gs_ampl_res}) is  
    represented  by dashed line. The ground state amplitude is
    normalised in a way that the smooth part equals unity for $t=0$.
    }
  \label{fig:gsamp}
\end{figure}

The results are presented in Fig.~\ref{fig:gsamp} for $N=20$
particles and $-1<t<1$. In the central part of the cloud  
the ground state amplitude is indeed well approximated by few 
terms in the expansion
(\ref{eq:gs_ampl_res}). Further improvement can be achieved using the
perturbation theory around each saddle point similarly to the
perturbation theory described in Section~\ref{sec:perturb} 
in ther case of circular geometry.
Close to the edges of the atomic cloud,
which in our units correspond to $t=\pm 2$, the  saddle point
approximation we used to derive (\ref{eq:gs_ampl_res})
breaks down since for $t$ approaching one of the  edges the saddle
points  (\ref{eq:xpm}) coalesce at $\pm i$ and the second derivative of
the action (\ref{eq:s2pm}) vanishes. In this case the analysis of the
higher order expansion of  action (\ref{eq:s2pm}) is required and is
beyond the scope of this work.

Thus we see that our method of calculation provides the exact large 
$N$ asymptotic  of the ground state amplitude. We extend it to 
the calculation of the one-body density matrix in the subsequent 
sections for the case of particles in harmonic potential and circular 
geometry.

\section{One-body density matrix in harmonic potential}
\label{sec:corr}
The one-body density matrix can be written using the definition
(\ref{eq:g1_def_0}) together with the expression (\ref{eq:gs}) for the
ground state function  
\begin{equation}
  \label{eq:g1_def}
  g_1 (t,t') = (N+1)
    \frac{S_N(N)}{S_{N+1 }(N)} e^{-\frac{N}{4}(t^2+t'^2)}\;
    Z\left(t,t'\right),
\end{equation}
where $Z(t,t')$ is   obtained from (\ref{eq:znnN}) according to the 
following rule:
\begin{equation}
  \label{eq:Ztt}
  Z(t,t') = \lim_{n\to 1/2} Z_{4n} (t,t')\equiv
  \lim_{n\to 1/2} Z_{4n} 
  (\underbrace{t,\ldots t}_{2n},\underbrace{t',\ldots t'}_{2n}) 
\end{equation}
It is again important to calculate $Z_{4n} (t,t')$ in the large $N$
limit before taking analytical continuation $n\to 1/2$, otherwise the
result will be just the one-body density matrix for one-dimensional fermions.

The duality transformation can be worked 
out in the case of two variables as explained in the 
Appendix~\ref{app:duality_der} and yields
\begin{eqnarray}
  \label{eq:Z2n_dual}
  Z_{4n} (t,t') &=& \frac{1}{S^2_{2n}(N)} 
  \int_{-\infty}^\infty d^{2n} x\,d^{2n}
  x'\,\Delta^2_{2n} (x)\Delta^2_{2n} (x') 
  \frac{\prod_{a,a'=1}^{2n} (x_a-x'_{a'})} 
  {\left[i\left(t-t'\right)\right]^{4n^2}} e^{-N\sum
  S(x_a,t)} e^{-N\sum S(x'_{a'},t')},
\end{eqnarray}
with the same definition (\ref{eq:sxt}) for  the effective action, $S(x,t)$.  

Before calculating the one-body density matrix by the saddle point method we
would like to remark that this correlation function depends on the
scaling of the distance $t-t'$ when $N$ goes to infinity. We
distinguish two limits: macroscopic, when $t,t'$ are of order 
of the cloud size and microscopic, or, more precisely,
 mesoscopic
limit, when $t,t'$ remains finite on the scale of mean inter-particle
distance $1/N\rho(t)$ equal to $\pi/N$ in the centre of the cloud. 
The second limit is called mesoscopic, since though $t,t'$ are small 
on the scale of the cloud size, we are interested in the asymptotic 
behaviour of correlations $t-t' \gg 1/N$, so there are still a very 
large number of particles between $t$ and $t'$.

In the macroscopic limit, the variables are changed in the following
way:
\begin{eqnarray}  
  x_a&=&x_-+\xi_a/\sqrt{N},\;\;\;\;a=1,\ldots,n+k
  \nonumber\\
  x_b&=&x_++\xi_b/\sqrt{N},\;\;\;\;b=n+k+1,\ldots,2n\nonumber\\
  x'_{a'}&=&x'_++\xi'_{a'}/\sqrt{N},\;\;\;\;a'=1,\ldots, n+k'
  \nonumber\\
  x'_{b'}&=&x'_-+\xi'_{b'}/\sqrt{N},\;\;\;\;b'=n+k'+1,\ldots,2n
  \label{eq:x_2_exp}
\end{eqnarray}
where $x_\pm=x_\pm (t)$ and $x'_\pm=x_\pm (t')$ defined in
Eq.~(\ref{eq:xpm}). The stationary value of the action and fluctuation
integrals are done exactly as in the last section.  The only new
factor is the double product in the integral, calculated at the saddle
point $(k,k')$ with the result
\begin{eqnarray}
  \label{eq:double_prod_res}
  \prod_{a=1}^{2n}\prod_{a'=1}^{2n} (x_a-x'_{a'}) = i^{4n^2}\, i^{2n(k+k')}
  \left|t-t'\right|^{n^2}
  \left[\frac{\cos^2\frac{\phi+\phi'}{2}}
  {\sin^2\frac{\phi-\phi'}{2}}\right]^{kk'}
  e^{-2in\left(k\phi-k'\phi'\right)}
\end{eqnarray}
This factor represents the interaction between saddle points in $x_a$
and $x'_b$ and as we shall see it is crucial for obtaining the correct
expression for $g_1$ in the mesoscopic limit.
Combining this factor with each $(k,k')$ saddle point contribution and
multiplying by the number of ways to distribute $x$, $x'$ among different
stationary values we get the integral (\ref{eq:Z2n_dual}) as a double
sum
\begin{eqnarray}
  &&Z_{4n} (t,t') = A^2_n \left(2\pi
    N\right)^{2n^2}  e^{-Nn \left(2-\left(t^2+{t'}^2\right)/2\right)} 
  \times\frac{\left[\rho (t)\rho
    (t')\right]^{n^2}} {\left|
    t-t'\right|^{2n^2}}\nonumber\\
    &\times&\sum_{k=-\infty}^{\infty}\sum_{k'=-\infty}^{\infty}    
    \left|\frac{\cos\frac{\phi+\phi'}{2}} {\sin
        \frac{\phi-\phi'}{2}}\right|^{2kk'} 
    \frac{(-1)^{nk}D^{(n)}_k} {\Big[8N\pi^3\rho^3(t)\Big]^{k^2}}
    \;e^{-iNk\Theta-4ink\phi}
    \frac{(-1)^{nk'}D^{(n)}_{k'}} {\Big[8N\pi^3\rho^3(t)\Big]^{{k'}^2}}
    \;e^{iNk'\Theta'+4ink'\phi'} ,\nonumber\\
  \label{eq:Z2n_dual_saddle1}
\end{eqnarray}
where  the summations are extended to infinity, relying on the factors
$D^{(n)}_k$ which cut off the finite number of terms in the sum.

We see immediately that again the most replica  
asymmetric saddle point $(k,k')=0$ in each
set of variables $x_a$ and $x'_{a'}$ provides the dominant 
smooth contribution, which after analytic continuation $n\to 1/2$ 
and normalisation given in (\ref{eq:g1_def})  yields 
the macroscopic one-body density matrix
\begin{equation}
  \label{eq:g1_macro}
  \bar{g}_1 (t,t') = \rho_\infty \left(\frac{N}{\pi}\right)^\frac{1}{2} 
  \;\frac{\left[\rho (t)\rho (t')\right]^\frac{1}{4}} {\left|
  t-t'\right|^\frac{1}{2}} .
\end{equation}
This result is identical to that of Ref.~\cite{Forrester2003} and
agrees completely  with the functional form deduced in \cite{Gangardt2003}.
The finite size correction to (\ref{eq:g1_macro}) are obtained by
taking $n=1/2$ in each terms summed up  in (\ref{eq:Z2n_dual_saddle1}) 
with the result
\begin{equation}
  \label{eq:g1_osc_h}
  \frac{g_1(t,t')}{\bar{g}_1(t,t')}-1=
  \sum_{(k,k')\neq (0,0)}    
    (-1)^{(k+k')/2}D^{(1/2)}_k D^{(1/2)}_{k'}
    \left|\frac{\cos\frac{\phi+\phi'}{2}} {\sin
        \frac{\phi-\phi'}{2}}\right|^{2kk'} 
    \frac{\;e^{-iNk\Theta-2ik\phi+iNk'\Theta'+2ik'\phi'}}
  {\Big[8N\pi^3\rho^3(t)\Big]^{k^2}\Big[8N\pi^3\rho^3(t)\Big]^{{k'}^2}}    
\end{equation}

Going to the mesoscopic limit we focus on the centre of the potential
$t+t'=0$ and define the scaling variable $x$ such that $t-t' = x/N$.
The factor in (\ref{eq:Z2n_dual_saddle1}) becomes
\begin{equation}
  \label{eq:sing_factor}
  \left|\frac{\cos\frac{\phi+\phi'}{2}} {\sin
          \frac{\phi-\phi'}{2}}\right|^{2kk'} \simeq
          \left(\frac{4N}{x}\right)^{2kk'}.
\end{equation}
Adding powers of $N$ in this expression with those in
(\ref{eq:Z2n_dual_saddle1}) we see that the diagonal elements $k=k'$
in the sum are of order 1, while the off-diagonal elements are at
least smaller by factor $1/N$. Therefore we anticipate that
the mesoscopic limit is given by the
diagonal part of the sum (\ref{eq:Z2n_dual_saddle1}).

In principle, one should reconsider the asymptotic expansion
(\ref{eq:Z2n_dual_saddle1}) in the mesoscopic limit, since in this
case the distance between saddle point is of order $1/N$, which can
change the contribution of the fluctuations. These calculations can be
done straightforwardly along the lines of 
Ref.~\cite{KamenevMezard1999}. Somewhat surprisingly, the conclusion 
is that the
asymptotic limit $N\to\infty$, and the mesoscopic limit $t-t'\to 0$
commute, so one can change the variables to $x$ directly
in the sum (\ref{eq:Z2n_dual_saddle1}). Apart from the factor
(\ref{eq:sing_factor}) the other factors simplify as $\rho (t)=\rho
(t')\to 1/\pi$, $\Theta-\Theta'=4\phi-4\phi'=2 x/N$ and one gets
\begin{equation}
  \label{eq:Z2n_meso}
  Z_{4n} (x) = A^2_n e^{-Nn\left(2-\left(t^2+{t'}^2\right)/2\right)}
     \left(\frac{\sqrt{2}N}{|x|^{1/2}}\right)^{4n^2}
    \left[1+2\sum_{k=1}^\infty (-1)^{2nk}\left[D^{(n)}_k\right]^2
    \frac{\cos 2kx}{(2x)^{2k^2}} \right] .
\end{equation}
Normalising and putting $n = 1/2$ one recovers the dominant term:
\begin{equation}
  \label{eq:g1_meso_res}
  \bar{g}_1 (x) = \left(\frac{N}{\pi}\right) \frac{\rho_\infty}{x^{1/2}}
\end{equation}
which is identical to the expression (\ref{eq:g1_smooth}) of 
Vaidya and Tracy if one identifies $k_F =\pi N\rho(0) = N$ and 
normalises to the density $N\rho(0)=N/\pi$ in the centre of the
cloud. By the same procedure
we  obtain  the oscillatory corrections
\begin{equation}
  \label{eq:g1_meso_res_osc}
  \frac{g_1 (x)}{\bar{g}_1 (x)} -1
    = 2\sum_{k=1}^\infty (-1)^{k}\left[D^{(1/2)}_k\right]^2
    \frac{\cos 2kx}{(2x)^{2k^2}}.
\end{equation}
Using the explicit value  $D^{(1/2)}_1=1/2$ we see that 
the $k=1$ term is identical to the 
the leading term (\ref{eq:vaidya_tracy})   of Vaidya and Tracy
expansion \cite{VaidyaTracy1979} apart from the sign. To our knowledge
it is the first analytical justification of the sign inconsistency in
the Vaidya and Tracy asymptotic expression first noticed in
\cite{Creamer1981} by using numerical solution of Painlev\'e equation.
The same sign change appears in the    thermodynamical limit of
circular geometry which we consider in the next section.

The expression (\ref{eq:g1_osc_h}) and (\ref{eq:g1_meso_res_osc})
together with the expressions (\ref{eq:gs_ampl_res}) of the 
coefficients $D^{(1/2)}_k$ provide only the  leading non-perturbative
 contribution of  each saddle point and do not
contain the sub-leading terms, which arise, for example, from the
deviation of the action from its second order Taylor  expansion near 
saddle point or additional terms in the large $N$ expansion of the
double product (\ref{eq:sing_factor}). These terms can be treated by a
perturbation theory near each saddle point and we show in
Section~\ref{sec:perturb}  how such
perturbation theory can be constructed for the case of circular
geometry.

\section{One-body density matrix in circular geometry}
\label{sec:circular_geometry}

Our method is also able to give the one-body density matrix in the
circular geometry, \textit{i.e.} for  $N+1$ particles on a ring
of length $L$ with periodic boundary conditions.  The one-body density
matrix is given by the expression (\ref{eq:circ_g1_def}).  
The corrections to the leading smooth term (\ref{eq:circ_smooth})
in the long-distance  expansion of the one-body density matrix
remained unknown to the best of our knowledge 
(see discussion in \cite{Forrester2003}).
We now show how they can be obtained with our method. Let us consider
the quantity (\ref{eq:average}) rewritten explicitly as an
$N$-dimensional integral
\begin{equation}
  \label{eq:circ_Zn_def}
  Z_{2n} (t) = \frac{1}{M_N(2n,2n,1)} \int_{-1/2}^{1/2} d^N \theta
  \left|\Delta_N \left(e^{2\pi i \theta}\right)\right|^2 \prod_{l=1}^N
  \left|1+e^{2\pi i \theta_l}\right|^{2n} \left|t+e^{2\pi i
  \theta_l}\right|^{2n},
\end{equation}
where $t=\exp(2\pi i \alpha)$ and we have normalised $Z_{2n}(1) = 1$,
introducing the normalisation constant $M_N(2n,2n,1)$. Its value  is given by
Morris integral of Random Matrix Theory \cite{ForresterWeb}, but its precise
value is not needed. Taking  $n\to 1/2$ in $Z_{2n}$  gives
the density matrix normalised to the density $g_1/n$.
Now we write
\begin{equation}
  \label{eq:abs_phase}
  \left|1+e^{2\pi i \theta_l}\right|^{2n} = e^{- 2\pi n i \theta_l}
  \left(1+e^{2\pi i \theta_l}\right)^{2n}
\end{equation}
and change the integration variables $\theta_l\to \theta+\alpha$ to
obtain
\begin{equation}
  \label{eq:circ_Zn_1}
  Z_{2n} (t) = \frac{t^{-Nn}}{M_N(2n,2n,1)} \int_{-1/2}^{1/2} d^N \theta
  \left|\Delta_N \left(e^{2\pi i \theta}\right)\right|^2 \prod_{l=1}^N
  e^{-2\pi n i\theta_l}\left|1+e^{2\pi i \theta_l}\right|^{2n} \left(1+t
  e^{2\pi i \theta_l}\right)^{2n} .
\end{equation}
One observes that the transformation from (\ref{eq:circ_Zn_def}) to
the last expression is possible only when $n$ is  integer,
otherwise the change of variables is not permitted due to
discontinuity  of the phase in (\ref{eq:abs_phase}). It is
parallel to the representation (\ref{eq:repr}) we used for the
harmonic confinement. As in the
last section we proceed with the representation (\ref{eq:circ_Zn_1}) 
and assume that it remains  valid for any $n$. 

The integral (\ref{eq:circ_Zn_1}) has 
a remarkable dual representation (see eq. (3.41) in \cite{Duality} )
by an integral over $n$ variables
\begin{equation}
  \label{eq:circ_duality}
  Z_{2n} (t) = \frac{t^{-Nn}}{S_{2n}(0,0,1)} \int_0^1 d^{2n} x\, \Delta_{2n}^2
  (x) \prod_{a=1}^{2n} (1-(1-t)x_a)^N,
\end{equation}
where the normalisation constant is given by Selberg integral
\begin{equation}
  \label{eq:circ_selberg}
  S_{2n}(0,0,1) = \int_0^1 d^{2n} x \,\Delta_{2n}^2 (x) = \prod_{a=1}^{2n}
  \frac{\Gamma^2(a) \Gamma(1+a) } {\Gamma(2n+a)}
\end{equation}
In the right hand side of (\ref{eq:circ_duality}) the number of
particles $N$ appears only as a parameter. This representation is a
direct analogy of (\ref{eq:Zn_dual}) and (\ref{eq:Z2n_dual}) and
allows us to obtain the asymptotic expression for $Z_{2n}$ suitable for
analytic continuation in $n$. In the large $N$ limit the integrand in
(\ref{eq:circ_duality}) oscillates rapidly and the main contribution
comes from the endpoints $x_\pm=1,0$ which are the only stationary
points of the phase. We change variables near each endpoint
 \begin{eqnarray}
   \label{eq:circ_chang_var}
   x_a&=&x_-+\frac{\xi_a}{N(1-t)}
,\qquad\qquad\qquad a=1,\ldots,l \nonumber\\
   x_b&=&x_+-\frac{\xi_b}{N(1-t^{-1})}
   ,\qquad\qquad b=l+1,\ldots,2n ,
 \end{eqnarray}
The integrand in (\ref{eq:circ_duality}) simplifies in the large $N$ limit:
\begin{equation}
  (1-(1-t)x_a)^N\simeq \left\{
    \begin{array}{ll}
      e^{-\xi_a}, & a=1,\ldots,l\\ 
      t^N e^{-\xi_a}, & a=l+1,\ldots, 2n
    \end{array}\right. 
  \label{eq:circ_spm}  
\end{equation}
and  the integration measure 
including the Vandermonde determinant is factorised as
\begin{equation}
  \label{eq:circ_vandermonde_fact}
  d^{2n} x\,\Delta^2_{2n} (x)
  = \left(\frac{1}{N(1-t)}\right)^{l^2}
  \left(\frac{1}{N(1-t^{-1})}\right)^{(2n-l)^2}
  d^{l}\xi_a\,\Delta^2_{l} (\xi_a) \;d^{2n-l}\xi_b\,\Delta^2_{2n-l} (\xi_b) 
\end{equation}
The  remaining integrals are calculated
using  Laguerre variant of the Selberg
formula
\begin{equation}
  \label{eq:lag_selberg}
 I_l (\lambda) = \int_0^\infty d^l \xi_a \Delta^2_l (\xi_a)
  \prod_{a=1}^m e^{-\lambda \xi_a} = \lambda^{-l^2} \prod_{a=1}^{l}
  \Gamma(a)\Gamma(1+a)
\end{equation}
Multiplying the contribution of each saddle point by the number of ways to
distribute variables we obtain the asymptotics of the integral
(\ref{eq:circ_duality}) as a sum of $2n+1$ terms:
\begin{equation}
  \label{eq:circ_Zn_2}
  Z_{2n} (t) = \prod_{c=1}^{2n} \Gamma (2n+c) \; \sum_l
   (-1)^{2n(n-l)}
   \frac{\left[F^l_{2n}\right]^2 t^{(N+2n)(n-l)}}
      {{(2X)^{l^2+(2n-l)^2}}}, 
\end{equation}
where we have introduced $X=N\sin\pi\alpha$ and 
the factors $F^l_{2n}$ are defined in (\ref{eq:Anl}).
We note that due to the translation invariance the resulting 
expansion is given by simple sum over
saddle points and not a double sum 
as in the case (\ref{eq:Z2n_dual_saddle1})  of harmonic potential. 

Changing the summation variable to $k=l-n$ and factorising the
 amplitudes $F^l_{2n}$ according to the 
 definitions (\ref{eq:Ank}), (\ref{eq:Dnk}) we obtain finally
\begin{equation}
  \label{eq:circ_Zn_3}
  Z_{2n} (t) = \frac{A_n^2 \;\prod_{c=1}^{2n} \Gamma (2n+c)}
  {(2N |\sin \pi\alpha|)^{2n^2}} \left(1+2\sum_{k=1}^\infty (-1)^{2nk}
    \left[D^{(n)}_k\right]^2 \frac{\cos \left[2k\, \pi (N+2n) \alpha\right] }
    {(4N^2\sin^2\pi\alpha)^{k^2}}\right)
\end{equation}
Now we are in a position to take the limit $n\to 1/2$  which results in the
desired corrections to the smooth part~(\ref{eq:circ_smooth}) of the
one-body density matrix:
\begin{equation}
  \label{eq:circ_g1_res}
  \frac{g_1 (\alpha)}{\bar{g}_1(\alpha)} - 1 
  = 2\sum_{k=1}^\infty
  (-1)^{k} \left[D^{(1/2)}_k\right]^2 
  \frac{\cos 2 k\,\pi (N+1) \alpha}{(4 N^2 \sin^2 \pi\alpha)^{k^2}}
\end{equation}
In the thermodynamic limit $N \pi\alpha = k_F x$, when $N\to\infty$ we
recover the result (\ref{eq:g1_meso_res_osc}). Again we note the sign
difference between the $k=1$ term of (\ref{eq:circ_g1_res}) and that of 
Vaidya and Tracy (\ref{eq:vaidya_tracy}). As in the case of harmonic
confinement only the leading contribution of each saddle point is included
in the expansion (\ref{eq:circ_g1_res}). The circular geometry is
particularly suitable for discussing the perturbative corrections,
which are calculated in the next section.

\section{Perturbation theory}
\label{sec:perturb}

Up to now we have considered only the leading non-perturbative
contribution of each saddle point in the expansion
(\ref{eq:circ_Zn_3}) of $Z_{2n}$. To provide all the corrections to a
given order in $1/N$ we have to deal with the sub-leading terms by 
perturbation theory. This amounts to multiplying the contribution of each
saddle point in (\ref{eq:circ_Zn_2}) by  the following average
\begin{eqnarray}
  \label{eq:pert_replace}
      \Big\langle F(\xi,\xi')\Big\rangle\equiv
      \int_0^\infty\! d^{n+k}\xi d^{n-k}\xi'\, 
      \Delta_{n+k}^2 (\xi)\Delta_{n-k}^2(\xi')\, F(\xi,\xi')\, 
      e^{-\sum\xi_a-\sum\xi'_b} 
\end{eqnarray}
of the function 
\begin{eqnarray}  
  F(\xi_1,\ldots,\xi_{n+k};\xi'_1,\ldots,\xi_{n-k}) &=&
  \prod_{a=1}^{n+k}\prod_{b=1}^{n-k} 
  \left(1-\frac{\Lambda_{ab}}{2X}\right)^2 \nonumber\\
  &\times&\exp\left(-\sum_{a=1}^{n+k}
    \left(\frac{\xi_a^2}{2N}+\frac{\xi_a^3}{3N^2}+\ldots\right)\right)
  \exp\left(-\sum_{b=1}^{n-k}
    \left(\frac{{\xi'}_b^2}{2N}+\frac{{\xi'}_b^3}{3N^2}+\ldots\right)\right) ,
  \label{eq:pert_F}
\end{eqnarray}
where we have defined $\xi'_b\equiv\xi_{b+n+k}$ for
$b=1,\ldots,n-k$ and
\begin{equation}
   \label{eq:pert_Lambda}
\Lambda_{(ab)} = 
   \frac{i}{2}\left(e^{\pi i\alpha} \xi'_b-e^{-\pi
       i\alpha}\xi_a\right) .
\end{equation}
In (\ref{eq:pert_F}) the product comes from the neglected
terms in the factorisation of the Vandermonde determinant, while the
exponents represent the corrections to the leading term
(\ref{eq:circ_spm}) in the $1/N$ expansion of the  action.

The perturbation theory consists in expanding $F(\xi,\xi')$ up to the
desired order in $1/N$ (recall that $1/X=1/N\sin\pi\alpha$) and
averaging (\ref{eq:pert_replace}) term by term with the unperturbed
action of each saddle point $k$.  It follows from the very essence of our
method, reflected in the structure of
(\ref{eq:pert_replace}), that the sets of variables $\xi_a$ and
$\xi'_b$ are independent, so that their  averages factorise
\begin{equation}
  \label{eq:pert_fact}
  \Big\langle f(\xi) g(\xi')\Big\rangle = \Big\langle f(\xi)\Big\rangle_{n+k} 
  \Big\langle g(\xi')\Big\rangle_{n-k}, \qquad\qquad
  \Big\langle f(\xi)\Big\rangle_{m}\equiv
  \frac{1}{I_m(1)}\int_0^\infty\! d^n\xi\, \Delta_{m}^2 (\xi)\, f(\xi)\, 
      \prod_{a=1}^m e^{-\xi_a}
\end{equation}
where the subscript reflects the number of integration variables in
each factor. The remaining averages are performed using known results 
from the theory of Selberg integrals (see chapter 17 of the
Ref.~\cite{MehtaRandMatr}).  In the following  we shall need the following 
results: 
\begin{eqnarray}
  \Big\langle\xi_1\Big\rangle_m &=& m,\nonumber\\ \nonumber\\
  \Big\langle\xi^2_1\Big\rangle_m &=& 2m^2,\nonumber\\
  \Big\langle\xi_1\xi_2\Big\rangle_m &=& m(m-1),\nonumber\\\nonumber\\
  \Big\langle\xi^3_1\Big\rangle_m &=& m(5m^2+1),\qquad\qquad\qquad\qquad
  \hfill\Big\langle\xi^2_1\xi_2\Big\rangle_m = m(m-1)(2m-1),\nonumber\\
  \Big\langle\xi_1\xi_2\xi_3\Big\rangle_m &=& m(m-1)(m-2),\nonumber\\\nonumber\\
  \Big\langle\xi^4_1\Big\rangle_m &=& m(8m^3+15m^2-2m+3),\qquad 
  \Big\langle\xi^3_1\xi_2\Big\rangle_m =m(m-1)(5m^2-4m+3),\nonumber\\
  \Big\langle\xi^2_1\xi_2\xi_3\Big\rangle_m &=&2m(m-1)^2(m-2), \qquad\qquad\;\;\;
  \Big\langle\xi^2_1\xi^2_2\Big\rangle_m =m(m-1)(2m-1)^2,\nonumber\\
  \Big\langle\xi_1\xi_2\xi_3\xi_4\Big\rangle_m &=&m(m-1)(m-2)(m-3)  
  \label{eq:pert_average}
\end{eqnarray}

In the first order we have
\begin{eqnarray}
  & &\frac{2}{X}\sum_{ab} \Big\langle\Lambda_{ab}\Big\rangle
  -\frac{1}{2N}\sum_a \Big\langle\xi^2_a\Big\rangle
  -\frac{1}{2N}\sum_b \Big\langle{\xi'}^2_b\Big\rangle
  = \nonumber \\
  &=& \frac{1}{X}\sum_{a=1}^{n+k}\sum_{b=1}^{n-k}
  \left[e^{\pi i\alpha} \Big\langle \xi'_b\Big\rangle_{n-k} -
    e^{-\pi i\alpha} \Big\langle \xi_a\Big\rangle_{n+k}\right] 
  -\frac{1}{2N}\sum_{a=1}^{n+k} \Big\langle\xi^2_a\Big\rangle_{n+k}
  -\frac{1}{2N}\sum_{b=1}^{n-k} \Big\langle{\xi'}^2_b\Big\rangle_{n-k}
  \nonumber\\
  &=&  
  \frac{i}{X} (n+k)(n-k)\left[ (n-k) e^{\pi i\alpha}-(n+k) e^{-\pi
      i\alpha}\right] - \frac{1}{N}\left[(n+k)^3+(n-k)^3\right]
  \label{eq:pert_first}
\end{eqnarray}
where we have used the fact that the averages are independent of the
index $a$ or $b$ and the corresponding sums yield a factor $n+k$ or 
$n-k$ respectively.  The calculation of the next order correction to
this term is performed similarly to the first order. One only has to
pay attention to the appearance of identical indexes in the sums. For
instance we have a sum of diagonal terms:
\begin{eqnarray}
  \label{eq:pert_second_diag}
  \frac{1}{X^2}\sum_{ab} \Big\langle\Lambda_{ab}^2\Big\rangle &=& 
  -\frac{(n+k)(n-k)}{4X^2}\left[ e^{2\pi
      i\alpha}\Big\langle{\xi'_1}^2\Big\rangle_{n-k}+
    e^{-2\pi i\alpha}\Big\langle\xi_1^2\Big\rangle_{n+k}-
    2\Big\langle\xi\Big\rangle_{n+k}\Big\langle\xi'\Big\rangle_{n-k}\right]
\end{eqnarray}
and non-diagonal ones:
\begin{eqnarray}
  \frac{2}{X^2}
  \sum_{ab\neq a'b'}\Big\langle\Lambda_{ab}\Lambda_{a'b'}\Big\rangle &=& 
   -\frac{(n+k)(n-k)}{2X^2}\Bigg(e^{2\pi i\alpha}
  \left[(n+k-1)\Big\langle{\xi'_1}^2\Big\rangle_{n-k}+(n+k)(n-k-1)
     \Big\langle\xi'_1\xi'_2\Big\rangle_{n-k}\right]\nonumber\\
   &+&e^{-2\pi i\alpha}\left[(n-k-1)\Big\langle\xi_1^2\Big\rangle_{n+k}
     +(n-k)(n+k-1)\Big\langle\xi_1\xi_2\Big\rangle_{n+k}\right]\nonumber\\ 
   &+&2\left((n+k)(n-k)-1\right)\Big\langle\xi_1\Big\rangle_{n+k}
   \Big\langle\xi'_1\Big\rangle_{n-k}\Bigg) 
  \label{eq:pert_second_non_diag}
\end{eqnarray}

Restricting ourselves to the terms up to the second order in 
perturbation theory  we see that only the case $k=0$ has to be
considered, since the leading contribution of 
$k\neq 0$ saddle points is already at least of  second order in $1/X$.
In this case
the first order terms (\ref{eq:pert_first}) simplify to 
$-4n^3/N$, while calculating the contribution of all the second order
terms in (\ref{eq:pert_F}), including that of (\ref{eq:pert_second_diag}) and 
(\ref{eq:pert_second_non_diag}) yields
 \begin{equation}
  \label{eq:pert_product}
 \Big\langle F(\xi,\xi')\Big\rangle =
  -\frac{n^4}{2X^2}+\frac{n^2}{N^2}\left(8n^4-n^3+\frac{23}{3}n^2-2n+\frac{1}{3}\right)  
\end{equation}
In every order in perturbation theory the coefficient is given by a
polynomial in $n$, so its analytical continuation to $n=1/2$ is
straightforward. A peculiar feature, familiar in the replica method,
of such perturbation theory of
this $n=1/2$ component field is that averages of positive 
quantities $\xi$, $\xi'$ vanish or become negative.

\begin{figure}[htbp]
  \centering
  \psfrag{c}{$C$}
  \psfrag{x}{$\alpha$}
  \psfrag{g}{$g_1(\alpha)/\bar{g}_1(\alpha)-1$}    
  \includegraphics[width=14cm]{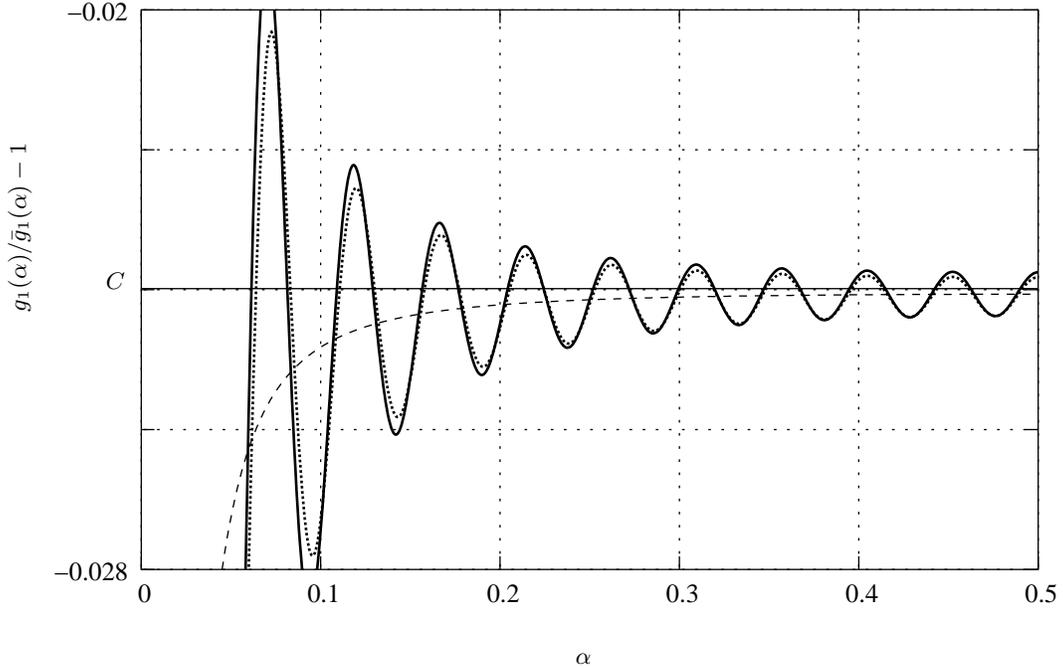}
  \caption{Comparison between asymptotic expression
    (\ref{eq:pert_res1}) for the one-body density matrix (solid line) 
    of $N=20$ particles in circular geometry
    and exact results (dotted line) based on numerical 
    evaluation of Toeplitz determinant. The correction due to
    the finite number of particles $C=-1/2N+13/32N^2=-0.023984375$ is
    shown and the dashed line represents the mean contribution of the 
    oscillatory terms. 
    }
  \label{fig:comparison}
\end{figure}

Adding up the second order contribution (\ref{eq:pert_product})
with that of first order (\ref{eq:pert_first}) and setting $n=1/2$  
we obtain the factor 
multiplying the contribution of the $k=0$ saddle point in the expansion
(\ref{eq:circ_Zn_3}). Combining it with the leading contribution of
the $k=\pm 1$ saddle point we get the finite size correction up to the
second order: 
\begin{equation}
  \label{eq:pert_res1}
  g_1 (\alpha)=  \frac{N\rho_\infty}{|N\sin\pi\alpha|^{1/2}}
    \left[1-\frac{1}{2N}+\frac{13}{32N^2}
      -\frac{1}{32N^2\sin^2\pi\alpha}
        -\frac{\cos 2\pi (N+1)\alpha}{8N^2\sin^2\pi\alpha}\right]
\end{equation}
We have compared this result with the exact calculation
based on a numerical evaluation of the Toeplitz determinant representation 
\cite{Lenard1964,Forrester2003} of $g_1 (\alpha)$. The result of
the comparison are presented in Fig.~\ref{fig:comparison} for $N=20$
particles. The agreement of the  two expressions is remarkable. 

\begin{figure}[htbp]
  \centering
  \psfrag{x}{$k_F x$}
  \psfrag{a}{a)}
  \psfrag{b}{b)}
  \psfrag{c}{c)}
  \psfrag{d}{d)}
  \includegraphics[width=16cm]{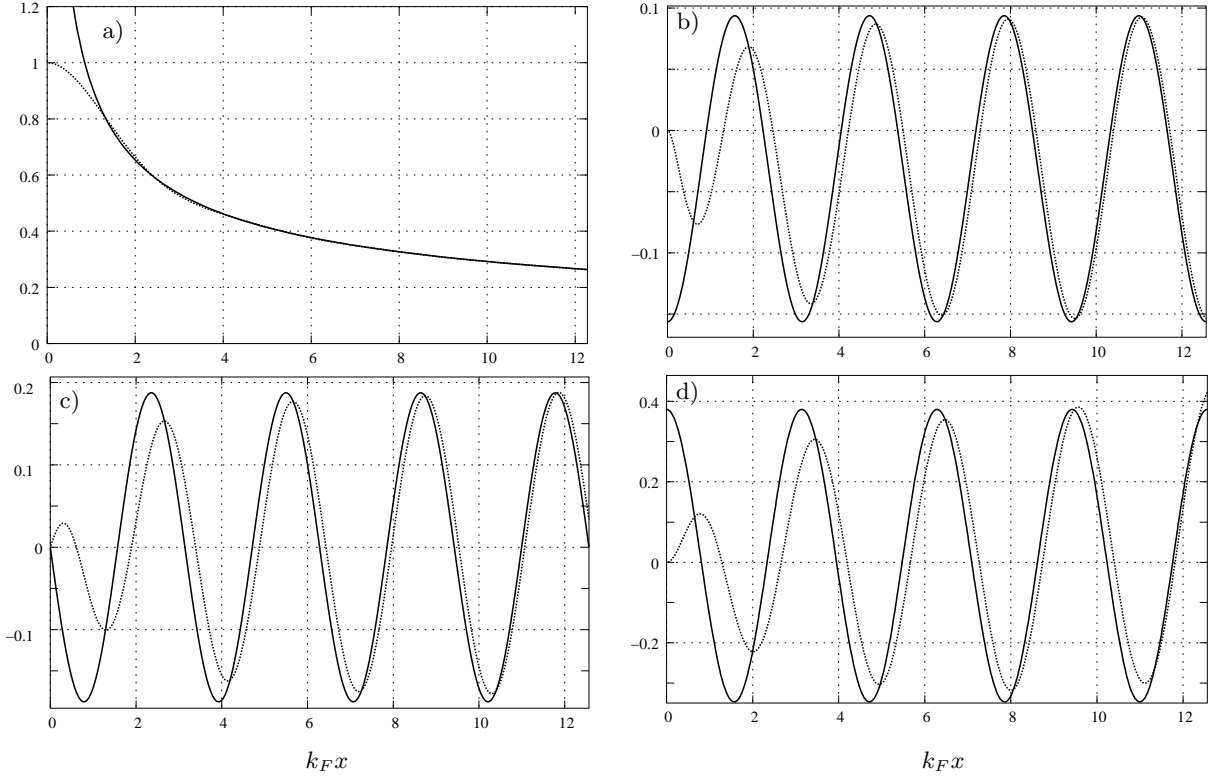}
  \caption{Comparison between different orders in $1/k_F x$ in 
    the asymptotic expansion 
    (\ref{eq:pert_res_therm})
    and results  based on numerical
    calculation of the one-body density matrix
    $g_1 (k_F x)$ on the lattice with filling factor $\nu=0.005$.
    a) Dotted line: the Toeplitz determinant evaluation of 
    $g_1(k_F x)$, solid line:
    $\bar{g}_1 (k_F x)=\rho_\infty/|k_F x|^{1/2}$.
    b) Dotted line: 
    $G^{(1)}(k_F x)=(k_F x)^2(g_1(k_F x)/\bar{g_1}(k_F x) -1)$, 
    solid line: $G^{(1)}_{th}=-1/32-\cos 2k_F x/8$.
    c) Dotted line: 
    $G^{(2)}(k_F x)=k_F x(G^{(1)}(k_F x)-G^{(1)}_{th}(k_F x))$, 
    solid line: $G^{(2)}_{th}(k_F x)=-(3/16)\sin 2k_F x$.
    d) Dotted line: 
    $G^{(3)}(k_F x)=k_F x(G^{(2)}(k_F x)-G^{(2)}_{th}(k_F x))$, 
    solid line: $G^{(3)}(k_F x)=33/2048+(93/256)\cos 2k_F x$.
  }
  \label{fig:fit_latt}
\end{figure}

Higher order corrections can be calculated in similar way. However the
number of averages proliferates quickly and the computations become
very tedious. The simplification occurs in  the thermodynamic limit,
where terms  proportional to just inverse powers of $N$ in (\ref{eq:pert_F})
can be neglected. In this limit only the double product
contributes in (\ref{eq:pert_F}) and the exponential factors in 
(\ref{eq:pert_Lambda}) can be set to one from the very beginning.
In this case we were able to
proceed up to the terms of order $1/X^4$, reproducing the results of
the asymptotic expansion of Vaidya and Tracy.    
Identifying in the thermodynamic limit
$X=N\sin\pi\alpha=N\pi\alpha=k_F x$  we get 
\begin{equation}
  \label{eq:pert_res_therm}
  g_1(k_F x) = \frac{\rho_\infty} {|k_F x|^{1/2}} \left[1-\frac{1}{32}
    \frac{1}{(k_F x)^2} -\frac{1}{8}\frac{\cos(2k_F x)}{(k_F x)^2}
      -\frac{3}{16}\frac{\sin (2k_F x)}{(k_F x)^3}+
   \frac{33}{2048}\frac{1}{(k_F x)^4}
    +\frac{93}{256}\frac{\cos (2k_F x)}{(k_F x)^4}\right]
\end{equation}
To compare this result with the numerics  
we use the representation for $g_1$ given in the work of 
Schultz \cite{Schultz1963}  as a continuum limit of the one-body density
matrix on a lattice which was a
starting point for Vaidya and Tracy calculations \cite{VaidyaTracy1979}.
It allows for  the direct study of $g_1$ in the thermodynamical limit 
if the distance on the lattice $d$ is related to the dimensionless 
distance in (\ref{eq:pert_res_therm}) as $\pi \nu d=k_F x$. 
The filling factor  $\nu$ approaches zero while  $k_F x$ is held 
fixed in order to reproduce the continuous limit. The results of the
comparison are shown in Fig.~\ref{fig:fit_latt} and the agreement is
good order by order. In particular, our result
(\ref{eq:pert_res_therm}) implies  that the sign of all the
trigonometric terms in Vaidya and Tracy expansion should be reversed
as in (\ref{eq:pert_res_therm}), while the values of numerical coefficients
remain unaffected.  

\section{Conclusions}
\label{sec:disc}

The main results of our calculations are  expressions
for the one-body density matrix accounting for finite size effects
in  the cases of  harmonic confinement  and circular
geometry. These
expressions consist of a sum of the leading smooth term and sub-leading
oscillating terms. These terms involve universal amplitudes $D^{(1/2)}_k$ 
given explicitly  by Eq.~(\ref{eq:gs_ampl_res}).  
The oscillations occur with periods $2k_F$, $4k_F$,
\ldots and are similar to  Friedel oscillations in fermionic
systems.  Their physical origin lies in particle correlations
on scales of mean inter-particle separation and they
represent a generic property of all strongly correlated one-dimensional
systems. For fermions their existence is due to the Pauli principle
and is not restricted to 1D, while for bosons the Friedel oscillations
are due to the effective Pauli principle induced by strong
interactions, a situation that is  possible only in the
one-dimensional  world.
Our results are in agreement with the general structure of the asymptotic 
expansion of $g_1$ given in \cite{VaidyaTracy1979} and with
Haldane's hydrodynamic approach \cite{Haldane1981} or conformal field
theory \cite{TsvelikBook}.  

Using semiclassical methods, such as local density approximation,
 this  universal  behaviour  of correlations can be extended  for a 
wide class of  external  confining potentials with  sufficiently smooth behaviour. 
Our calculations provide \emph{ab initio}
justification of the local density approximation in the bulk of
bosonic cloud in a harmonic trap which is important to current
experiments with cold atoms. For this system our method is capable to 
consider the edges  of thermodynamical density profile, thus going beyond the local 
density approximation.  Another system of experimental 
interest is the lattice model of impenetrable bosons, produced
recently  \cite{Paredes2004}  by confining atoms in optical lattices. 
Given an exact expression for the one-body density
matrix in terms of Toeplitz determinants \cite{Schultz1963,Lenard1964}
it is highly desirable to obtain its asymptotics without going to the
continuum limit, thus revealing the interplay between the short-range
correlations and the effects of lattice.

The results were obtained using a novel modification of the replica method
which in the context of exactly solvable models consists of 
alternative representation of correlation functions. 
The phenomenon of replica symmetry breaking serves here as a tool to
single out the bosonic branch of the one-body density matrix out of
various possible analytic continuations in the
replica index $n$. It can be considered as a fresh insight in quantum
statistics in one dimensions \cite{FracStat}, a question to explore in
the future. The obvious candidate for this study is the Calogero-Sutherland model.
Indeed, the ground state wave-function of  the Calogero-Sutherland
model is proportional to the Vandermonde determinant
(\ref{eq:vandermonde}) taken to a power $\lambda$, which characterises
the statistical interactions between the particles
\cite{Ha1995}. Based on equivalence of Calogero-Sutherland models and
Random Matrices for particular values of statistical interaction
parameter $\lambda$,  the replica method also provides an efficient 
way to study correlations of spectral
determinants of random matrices directly related to  averages 
of the form  (\ref{eq:average}). These objects  play an important 
role in statistical physics, mathematical physics and  modern 
combinatorics (see \cite{BrezinHikami2000,Conrey2003} and references therein). 
A more distant, but certainly tempting perspective is the 
application of the Replica Method to other  integrable models and  
calculation of their correlation properties.

\acknowledgements

The author is grateful to Marc M\'ezard, Gora Shlyapnikov, Eugene Bogomolny
and Stephan Ouvry for fruitful discussions and to Vanja Dunjko 
for numerical results and encouraging communications which had an 
important impact on the present work. Discussions with Craig Tracy,
Alex Kamenev, Tim Garoni and Antonio Garc\'ia-Garc\'ia are kindly acknowledged. 
The work was supported by grants from ENS and CNRS.
LPTMS is mixed research unit No. 8626 of CNRS and Universit\'e Paris
Sud.  LKB is mixed research unit No. 8552 of CNRS, of 
Ecole Normale Sup\'erieure and of Universit\'e Pierre et Marie Curie and CNRS.

\appendix
 
\section{\label{app:duality_der} Derivation of the duality formula }

Using the obvious identity among the Vandermonde determinants
\begin{equation}
  \label{eq:ident_vandermond}
  \prod_{j=1}^N \prod_{a=1}^m (t_a-y_j) = \frac{\Delta_{N+m}
  (t_1,\ldots,t_m,y_1,\ldots,y_N)}{\Delta_m(t_1,\ldots,t_m)\Delta_N
  (y_1,\ldots, y_N)}
\end{equation}
the expression (\ref{eq:znnN}) is  represented in the second
quantisation
\begin{equation}
  \label{eq:Z=znnN_2nd}
  Z_m (t_1,\ldots,t_m) = \sqrt{\frac{S_{N+m}}{S_N}} e^{\frac{N}{4}\sum
  t_a^2} \frac{1}{\Delta_m (t)}\; {}_N \langle \psi(t_1) \psi(t_2)
  \ldots \psi (t_m)\rangle_{N+m},
\end{equation}
using the fermionic annihilation operators $\psi(x)$ acting between 
ground state of $N$ and $N+m$  fermions. One uses then 
Wick theorem to calculate this matrix element:
\begin{equation}
  \label{eq:wick}
  {}_N \langle \psi(t_1) \psi(t_2) \ldots \psi (t_m)\rangle_{N+m} =
  \det_{k,l} \left[\phi_{N+k-1} (t_l)\right].
\end{equation}
The resulting determinant in the right hand side of (\ref{eq:wick})
is constructed using  the one-particle wave functions
(\ref{eq:orbitals}), which have
an integral representation
\begin{equation}
  \label{eq:int_repr}
  \phi_k (t) = e^{-\frac{Nt^2}{4}}\sqrt{\frac{2^k N^{k+1}}{2\pi
  c_k}}\frac{1}{i^k} \int_{-\infty}^\infty dx\, x^k
  e^{-\frac{N}{2}(x-it)^2}.
\end{equation}
Representing in this way the one-particle wave functions in the
expansion of the determinant (\ref{eq:wick}) we have the following
result:
\begin{equation}
  \label{eq:ZnnN_x}
  Z_m (t_1,\ldots,t_m) = \left(\frac{N}{2\pi}\right)^\frac{m}{2}
  (-i)^{Nm} \int_{-\infty}^\infty d^m x\, e^{-\frac{N}{2}\sum
  \left(x_a-it_a\right)^2 } \frac{\Delta_m (x_1,\ldots,x_m)}{\Delta_m
  (it_1,\ldots,it_m)} \prod_{a=1}^m x_a ^N
\end{equation}

In order to calculate various correlation functions, such as density
matrix (\ref{eq:g1_def}) or ground state amplitude
(\ref{eq:gs_ampl_def}) we need to take the limit where
several variables $t_a$ become equal to each other. 
This limit is finite, despite the apparent
singularity in the last integral representation of $Z_m$. We
demonstrate this in the simpler case of ground state amplitude
obtained from (\ref{eq:zt_replica}).
Shift the variables $t_a=t+\eta_a$, $\eta_a\to 0$
and rewrite the expression (\ref{eq:ZnnN_x}) as
\begin{equation}
  \label{eq:ZnnN_x1}
  Z_m = \left(\frac{N}{2\pi}\right)^\frac{m}{2} (-i)^{Nm}
  \int_{-\infty}^\infty d^m x\,\Delta_m (x) \frac{e^{iN\sum
  x_a\eta_a}} {\Delta_m (i\eta)} \prod_{a=1}^{m} x_a^N
  e^{-\frac{N}{2}\left[\left(x_a-it\right)^2 -\eta_a^2\right] } ,
\end{equation}
Due to the
presence of totally antisymmetric function $\Delta_m(x)$ only the
totally antisymmetric part of $e^{iN\sum x_a\eta_a}$ survives the
integration. Using this and the fact  that
 \begin{equation}
   \label{eq:sing_lim}
   \lim_{\eta\to 0} \frac{\det_{a,b} e^{iNx_a\eta_b}}{\Delta_m(i\eta)}
  = \frac{N^{m(m-1)/2}}{\prod_{a=0}^m \Gamma (a+1)} \Delta_m (x)
 \end{equation}
we arrive at the dual representation (\ref{eq:Zn_dual}).

The case of two variables $t$, $t'$ is similar. One starts with the
expression (\ref{eq:znnN}) and shifts the variables
\begin{eqnarray*}
  t_a &=& t+\eta_a,\;\;\;\;\; a=1,\ldots,m/2 \\ t_b &=&
  t'+\eta_b,\;\;\;\;\; b=m/2+1,\ldots,m,
\end{eqnarray*}
where $\eta_a$ and $\eta_b$ go to zero independently.  One uses the
fact (\ref{eq:ident_vandermond}) that
\begin{eqnarray*}
  \Delta_{m} (it_1,\ldots,it_{m}) = \Delta_{m/2}
  (i\eta_a)\Delta_{m/2}(i\eta_b) \left[i\left(t-t'\right)\right]^{m^2/4}
\end{eqnarray*}
to the leading non-vanishing order in $\eta_a$, $\eta_b$ and rewrites
(\ref{eq:ZnnN_x}) as
\begin{eqnarray*}
  Z_{m} (t,t') &=& \left(\frac{N}{2\pi}\right)^{m/2} (-i)^{mN}
  \int_{-\infty}^\infty d^{m} x\, \frac{\Delta_{m}
  (x_1,\ldots,x_{m})} {\left[i\left(t-t'\right)\right]^{m^2/4}}
  \frac{e^{iN\sum x_a\eta_a}}{\Delta_n (i\eta_a)} \frac{e^{iN\sum
  x_b\eta_b}}{\Delta_{m/2} (i\eta_b)} \\ &\times& \prod_{a=1}^{m/2} x_a^N
  e^{-\frac{N}{2} \left[\left(x_a-it\right)^2 -\eta_a^2\right] }
  \prod_{b=1}^{m/2} x_b^N e^{-\frac{N}{2} \left[\left(x_b-it'\right)^2
  -\eta_b^2\right] } ,
\end{eqnarray*}
Anti-symmetrising the numerators  in each set of variables $x_a$ and
$x_b$ independently  and using
(\ref{eq:sing_lim}) one gets (\ref{eq:Z2n_dual}).

\section{Analytical continuation of $A_n$}
\label{app:analAn}

We use the following integral representation \cite{GradshteynRyzhik} 
for the logarithm of Euler's gamma function 
\begin{equation}
  \label{eq:log_gamma}
  \ln \Gamma (z) = \int_0^\infty \frac{dt}{t}
  \left(\frac{e^{-zt}-e^{-t}}{1-e^{-t}}+(z-1) e^{-t}\right)
\end{equation}
to represent the logarithm of $A_n$ as an integral
\begin{equation}
  \label{eq:logAn}
  \ln A_n=\sum_{a=1}^{n} \left(\ln \Gamma (a) -\ln \Gamma
  (2n+1-a)\right) =\int_0^\infty \frac{dt}{t} e^{-t} \left[
  \left(\frac{1-e^{-n t}}{1-e^{-t}}\right)^2
  -n^2\right],
\end{equation}
where we have summed finite geometric series under the integral.  The 
integral representation defines $A_n$  for any value of $n$. 
In particular for $n=1/2$ we get
\begin{equation}
  \label{eq:logA1}
  \ln A_{1/2}=\int_0^\infty \frac{dt}{t} e^{-t} \left[
  \left(\frac{1-e^{-\frac{t}{2}}}{1-e^{-t}}\right)^2
  -\frac{1}{4}\right]=\frac{1}{4}\int_0^\infty \frac{dt}{t}\,
  \frac{3e^t-8e^{t/2}+6-e^{-t}}{\left(e^t-1\right)^2}
\end{equation}
In order to calculate the last integral we regularise the divergence
at $t=0$ in the following way:
\begin{equation}
  \label{eq:logA1nu}
  \ln A_{1/2}=\lim_{\nu\to 0} C(\nu),\qquad\qquad
  C(\nu)=\frac{1}{4}\int_0^\infty dt\,t^{\nu-1}
  \frac{3e^t-8e^{t/2}+6-e^{-t}}{\left(e^t-1\right)^2} 
\end{equation}
and calculate the integral term by term using the formula
\begin{equation}
  \label{eq:integral}
  \int_0^\infty \frac{x^{\nu-1} e^{-\mu x} dx}{\left(e^x-1\right)^2}=
  \Gamma (\nu) \left[\zeta(\nu-1,\mu+2)-(\mu+1)\zeta
  (\nu,\mu+2)\right],
\end{equation}
where $\zeta(z,q)$ is the Riemann's Zeta function
\begin{equation}
  \label{eq:riemann_zeta}
  \zeta(z,q) =\sum_{n=0}^\infty \frac{1}{(q+n)^z},\;\;\;\;\;
  \zeta(z,0)=\zeta(z).
\end{equation}
The result of integration can be represented as
\begin{equation}
  \label{eq:logA1nu1}
  C(\nu) = 2\Gamma (\nu) \left[2(1-2^{\nu-2})\zeta (\nu-1) - (1-2^{\nu
  -1}) \zeta(\nu) -1\right]
\end{equation}
Taking the limit by the l'H\^opitale rule and using the fact that
$\zeta(0)=-1/2$, $\zeta(-1)=-1/12$, $\zeta' (0) =-\ln \sqrt{2\pi}$ we
arrive at
\begin{equation}
  \label{eq:logA1_res}
  \ln A_{1/2} = \lim_{\nu\to 0} C(\nu) = 3\zeta'(-1) +\frac{1}{12}\ln 2
  +\frac{1}{2} \ln \pi,
\end{equation}
which relates $A_{1/2}$ to Glaisher's constant
$A=\exp\left(1/12-\zeta'(-1)\right)$ as 
\begin{equation}
\rho_\infty=A^2_{1/2}/\sqrt{2} = \pi e^{1/2} 2^{-1/3} A^{-6}.
\end{equation}
An alternative method of analytical continuation to that described
above is to relate the constants $A_n$ to Barnes $G$
function \cite{ForresterWeb}. The definition (\ref{eq:Ank}) yields
\begin{equation}
  \label{eq:a_barnes}
  A_n = \frac{\prod_{a=1}^n \Gamma (a)}{\prod_{a=1}^n \Gamma(n+a)}=
  \frac{G(n+1)}{G(1)}\frac{G(n+1)}{G(2n+1)}= \frac{G^2(n+1)}{G(2n+1)},
\end{equation}
where we have used $G(1)=1$ and the functional relation
$G(z+1)=\Gamma(z) G(z)$. The analytical continuation leads to identity
$A_{1/2} = G^2 (3/2)$ .


\begin{thebibliography}{99}
\bibitem{rmrkDunjko} Similar observation has been communicated to the
  author by V.~Dunjko, who matched numerically  short-distance
  and long-distance asymptotics of the one-body density matrix.
\end{thebibliography}

\end{document}